\documentclass[aps,groupedaddress,nofootinbib]{revtex4-1}

\usepackage{graphicx,siunitx,color,amsmath,float}      
\usepackage{tikz}
\usepackage{subfigure}

\begin{document}
\title{Population Dynamics Model and Analysis for Bacteria Transformation and Conjugation}
\author         {J.J. Dong, J.D. Russo, K. Sampson}
\email          {jiajia.dong@bucknell.edu}
\affiliation    {Department of Physics and Astronomy, Bucknell University, Lewisburg, PA, USA 17837}
\date{\today}

\begin{abstract}
  We present a two-species population model in a well-mixed environment where the dynamics involves, in addition to birth and death, changes due to environmental factors and inter-species interactions. The novel dynamical components are motivated by two common mechanisms for developing antibiotic resistance in bacteria: plasmid {\it transformation}, where external genetic material in the form of a plasmid is transferred inside a host cell; and {\it conjugation} by which one cell transfers genetic material to another by direct cell-to-cell contact. Through analytical and numerical methods, we identify the effects of transformation and conjugation individually. With transformation only, the two-species system will evolve towards one species' extinction, or a stable co-existence in the long-time limit. With conjugation only, we discover interesting oscillations for the system. Further, we quantify the combined effects of transformation and conjugation, and chart the regimes of stable co-existence, a result with ecological implications.
  \end{abstract}

\maketitle

\section{Introduction}
\begin{figure}[h]
  \includegraphics[width=0.5\textwidth]{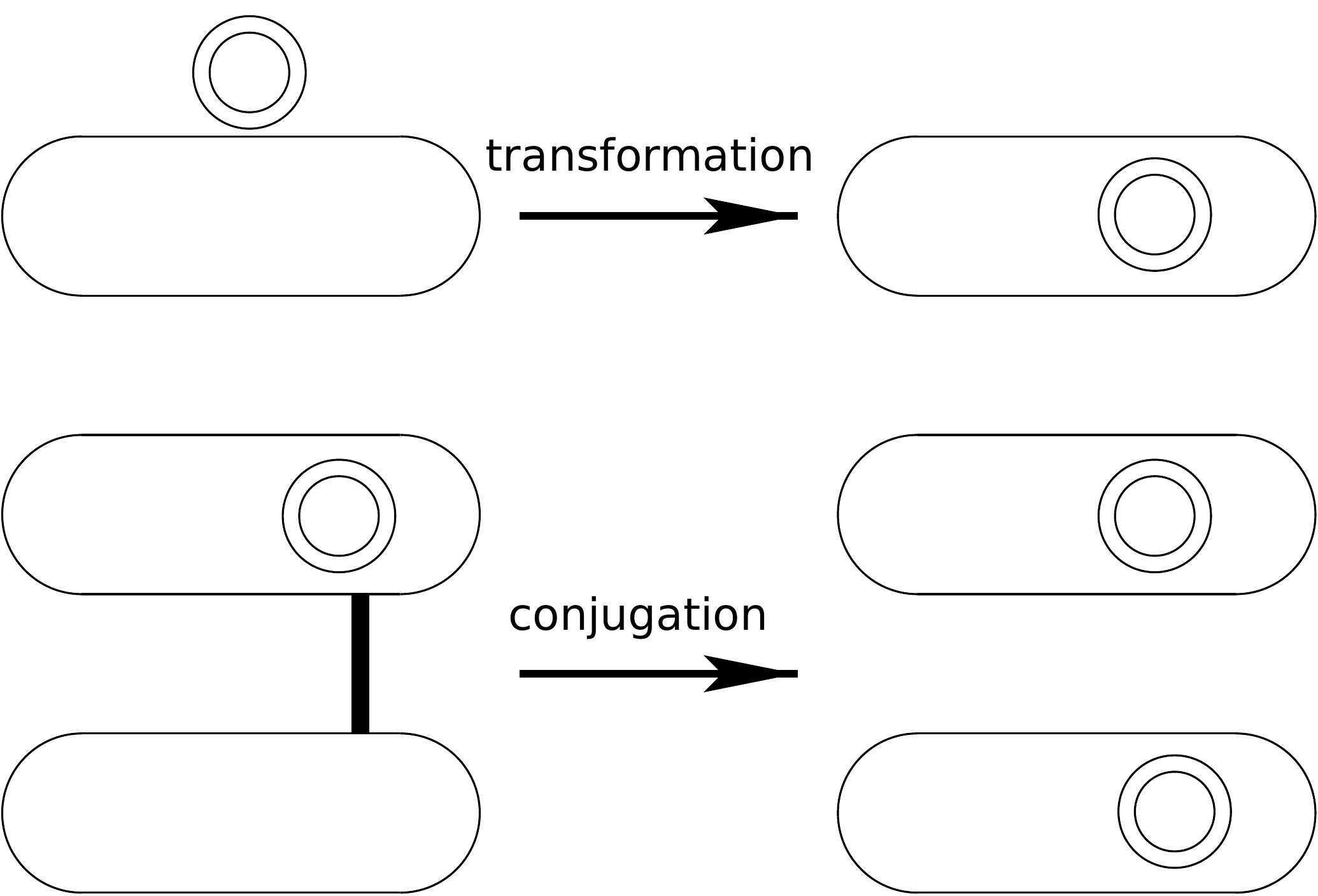}
  \caption{Schematics of the two resistance acquisition mechanisms: transformation (top) where the plasmid (circle) in the environment is incorporated into the bacterium, and conjugation (bottom) where a plasmid-carrying cell transfers the plasmid to a plasmid-free one through a bridge-like appendage (thick line) on the cell surface.} \label{fig:transf}
\end{figure}

The synthesis of large varieties of antibiotics is one of the most important medical inventions in human history. However, the emergence of antibiotic-resistant bacterial population, infamously known as the ``superbugs,'' poses an alarming challenge to public health 
authorities \cite{davies2010,WHO}, whose efforts in combating resistance are outpaced by the rapid development of new resistance in the bacteria cells \cite{barber49,wong02,livermore03}. Bacteria become resistant to antimicrobial agents as a result of exchanging genetic materials. 
Most of the bacterial DNA is contained in the chromosome, which provides the genetic identification for each cell. In addition to the chromosome, there are plasmids -- small circularized DNA molecules independent of chromosomes -- that have essential genes for 
plasmid functions and accessory genes \cite{maclean2015,PlasmidBiology}. It is the accessory genes that can confer antibiotic resistance to the host bacteria. There are two main mechanisms of horizontal gene transfer (HGT) through which an individual bacterium cell acquires antibiotic resistance from plasmids: 
\textit{transformation}, where a cell incorporates a plasmid from its surroundings, and \textit{conjugation}, where a plasmid-carrying cell transfers the plasmid (a single strand of the plasmid DNA) to a plasmid-free one by direct cell contact \cite{tatum1947}. A schematic of 
the two processes is shown in Fig.\ref{fig:transf}. The plasmids are transcribed and translated by the cellular machinery of the host cell and thus propagate in the population through cell division.

\begin{table*}[t]
  \centering
\begin{tabular}{|c |c l|l|l|}
\hline
~~index~~  &~~Reaction  & & ~~Description~~ &~~Typical values and references~~\\
\hline\hline
1&$S$ & $\stackrel{~~b_{S}~~}{\longrightarrow} 2S $& $S$ growth& $\sim0.5$ - $2 \si{hr}^{-1}$ \cite{monod1949,EcoliGrowth}\\\hline
2&$R$ & $\stackrel{~~b_{R}~~}{\longrightarrow} 2R$ \label{rxn:rbirth} & $R$ growth& $\sim(0.6$ - $1.0)\cdot b_S$ \cite{vogwill2015}\\\hline
3& $S+ P$&$\stackrel{~~\alpha ({P})~~}{\longrightarrow}R$& transformation& $ \sim10^{-4}$ - $10^{-6}\si{hr}^{-1}$ \cite{hanahan1983,lorenz1994} \\\hline
4&  $S+R$&$ \stackrel{~~\gamma~~}{\longrightarrow} 2R $\label{rxn:conjug} & conjugation& $\sim2.4$ - 7.6$\times10^{-3}\si{hr}^{-1}$  \cite{Korolev2014}\\\hline
5&$R $&$\stackrel{~~\delta~~}{\longrightarrow} 0 + P$ & $R$ lysis, releasing a copy of $P$& estimated in model  \\
   \hline
\end{tabular}    \label{rxn:death}
\caption{Interactions involved in the model and the reference reaction rates from experimental studies on various organisms.}\label{table:rxn}  
\end{table*}

Due to the short cell doubling cycle of bacteria, a small initial fraction of plasmid-carrying cells will be rapidly amplified in the population through division. There is an obvious advantage for plasmid-carrying cells in the presence of antibiotics. In the absence of antibiotics, however, the plasmid-carrying cells can still quickly outnumber the plasmid-free ones as evidenced by both laboratory 
and clinical findings \cite{ender04,hurdle04,andersson10}. The domination of either genotype depends on the respective doubling time as well as the rates of transformation and conjugation, which vary according to the plasmid concentration, the medium and the type of 
bacteria population. It is therefore crucial to quantitatively chart the parameter space and identify how the two populations compete under different conditions, which will bring further insights into the emergence of the resistance-dominant population.

The dynamics of biological systems has been extensively studied for decades using nonlinear differential equations such as the venerable Lotka-Volterra equation \cite{lotka20,volterra26}, agent-based modeling \cite{axelrod1997, DK}, and numerous tools developed in statistical physics \cite{domb1995,Jarzynski97}. Such interdisciplinary efforts have brought deeper insights into understanding biological systems of different scales, ranging from neuron networks to global pandemic spread, as well as expanding the understanding in generic complex systems. In this study, we take the biological processes of transformation and conjugation among bacteria cells as motivation, and construct a theoretical model for two growing bacterial populations of different genotypes inhabiting a well-mixed environment (e.g. a shaken flask): the plasmid-carrying cells that are antibiotic-resistant ($R$), and the plasmid-free ones ($S$) that are susceptible to antibiotics and can be converted to $R$ through either transformation or conjugation. The competition between the two sub-populations depends on the dynamics of the two HGT mechanisms and their respective doubling time, which reflects 
the plasmid carriage cost with varying benefit according to environmental conditions such as the level of antibiotics. By using a combination of deterministic calculation and numerical methods of solving coupled nonlinear differential equations, we examine the effects that these parameters have on the overall population in Sections \ref{sec:model} and \ref{sec:sim}. One of the main questions we attempt to answer in this study is with the interplay of transformation and conjugation in various growth regimes, how the parameter space is configured such that the overall system displays one species' extinction (or ``fixation'' by the other, which is commonly used when the changes are genotypic) or coexistence with one species dominating. We summarize and discuss the biological relevance in Section \ref{sec:sum}.

\section{Population Dynamics Model\label{sec:model}}
We formulate a model at the population level that incorporates growth, death and the mechanisms of HGT between the $S$ and $R$ populations in a well-mixed culture. Initially, the system is seeded with $S_0$ susceptible and $R_0$ resistant cells, which have growth rates $b_{S}$ and $b_{R}$. The microscopic growth rates of $S$ and $R$ are accessible through experiments by measuring the population doubling time $\mu_{S/R}$ of the two genotypes: $b_{S/R} =\ln 2\cdot(\mu_{S/R})^{-1}$. 

We focus on two HGT mechanisms through which cells develop antibiotic resistance: {\it transformation} and {\it conjugation}. When there is a large number of plasmids ($P$) in the environment, the transformation rate depends on the cell's innate ability to take up extracellular DNA. Therefore we choose a constant transformation rate $\alpha({P})=\alpha_0$ to model the {\it unlimited} plasmid supply or that the plasmid has a very high affinity to the cell. Accounting for the availability of free plasmids $P$ in the environment and the kinetics of plasmid uptake, we also study the scenario where the transformation rate depends  {\it linearly} on $P$ when the concentration of plasmids is low and transitions to the constant rate as $P$ increases, namely $\alpha({P})= \alpha_0P/(P+K_P)$. Here $K_P$  reflects the plasmid affinity to the cell and is the concentration of plasmids at which transformation occurs at the half-maximum rate. This form is analogous to the Michaelis-Menten kinetics\cite{PhysChem} in enzymatic activities. Other $P$-dependence scenarios can be studied with a similar approach.

During the process of {\it conjugation}, the double stranded plasmid DNA in the donor cell becomes two single strands, one of which
is transferred from the donor to the recipient through direct cell-to-cell contact mediated by a tubelike structure called ``pilus'' that pulls the recipient and perforates it \cite{Griffiths2000,Korolev2014}. Once the single strand of plasmid DNA is in the recipient, the DNA replication process 
restores the complementary strand and both the donor and the recipient possess the entire plasmid DNA. We model the conjugation process with constant rate $\gamma$ in the well-mixed population.

Finally, $R$ cells lyse\footnote{It is possible for $S$ cells to lyse as well. However the death of $S$ effectively reduces $b_S$, and we do not need introduce an additional parameter here.} with rate $\delta$. Unlike the phenotypical switching where a bacterium changes from a normal cell to a persistent cell in the presence of antibiotics with the same genetic content \cite{balaban2004}, a resistant cell dies when its cell wall breaks down. An $R$-cell is then removed from the population upon lysis and the plasmid it carries is released back to the environment. In the plasmid-limiting scenario, the recycling of plasmids proves to be significant in maintaining a resistant population. When there are antibiotics of various concentrations in the environment, the lysis rates of $S$ and $R$ populations lead to more interesting dynamics which we will not focus on in this article. 

The above reactions are summarized in Table \ref{table:rxn}. By using a combination of theoretical analysis for insights and numerical exploration for a wider range of parameters, we aim to provide a comprehensive picture of the roles transformation and conjugation play. To connect our theoretical model to experiments, we also include typical values of the parameters, such as $b_S, b_R, \alpha$ and $\gamma$, in our model. However, the cell lysis process is usually harder to ascertain and we will estimate the parameter $\delta$ in our model. More details and discussions will be included in Section \ref{sec:sim}.

The dynamics of the two populations, $S(t)$ and $R(t)$, can be described by the following set of equations after properly non-dimensionalize the parameters without loss of generality:
\begin{align}
\frac{dS}{dt} & = b_S S - \alpha
\left(P \right)S  - \gamma S R \label{eqn:sbirth} \\
\frac{dR}{dt} & = b_R  R + \alpha\left(P\right)
S  + \gamma S R   - \delta R\label{eqn:rbirth}
\end{align}

In addition, the plasmids in the environment vary due to the uptake by $S$ cells, as well as the release by $R$ cells upon their lysis:
\begin{align}
 \frac{dP}{dt} &  =  -\alpha \left( P\right) S + \delta R, ~~ \alpha(P)= \alpha_0 \dfrac{P}{P+K_P} \label{eqn:pfree}
\end{align}

To analyze the above set of coupled differential equations, we used a combination of theoretical analysis and numerical solvers. For the latter, we coded the equations using Python (v3.8) \cite{Python}. We used Euler's method \cite{butcher2008numerical} and the ODE solver  \verb,odtint,  from Scipy (v1.5) \cite{Scipy} to perform the numerical simulations, which yielded consistent results. The numerical results presented below are from Euler's method.

Before delving into a systematic analysis of the full model, we analyze the results of two limiting scenarios which provide guidance on the choice of parameters in the later simulations. 

\subsection{Transformation with constant $\alpha=\alpha_0$\label{sec:trans}}
We first consider the transformation-only case, namely $\gamma = 0$, and the transformation rate is a constant $\alpha_0$. This corresponds to the case where the environment concentration of plasmids is very high, or $K_P$ is very small. With the effective growth rate $\tilde{b}_S \equiv b_S-\alpha_0$, the dynamics of the $S$ population is an exponential function $S(t) = S_0 \exp{(\tilde{b}_St)}$. Similarly, the growth of the $R$ population involves an exponential term with $\tilde{b}_R \equiv b_R-\delta$ and the influx of the transformed $S$ cells:
\begin{eqnarray*}
R(t) &=& \dfrac{\alpha_0}{\tilde{b}_S-\tilde{b}_R}S(t)+\left[R_0-\dfrac{\alpha_0}{\tilde{b}_S-\tilde{b}_R}S_0\right] e^{\tilde{b}_Rt}\\
&=&R_0e^{\tilde{b}_Rt}+ \dfrac{\alpha_0}{\tilde{b}_S-\tilde{b}_R}S_0\left[e^{\tilde{b}_St}-e^{\tilde{b}_Rt}\right]
\end{eqnarray*}

The ratio between the two cell populations $f\equiv R/S$ is given by:
\begin{equation}
f(t) = \dfrac{\alpha_0}{\varepsilon}+\left[\frac{R_0}{S_0}-\dfrac{\alpha_0}{\varepsilon}\right]e^{-\varepsilon t}, 
\end{equation}
where $\varepsilon\equiv\tilde{b}_S-\tilde{b}_R$, the difference between the two effective growth rates. 

\begin{figure}[H]
\includegraphics[width=0.45\textwidth]{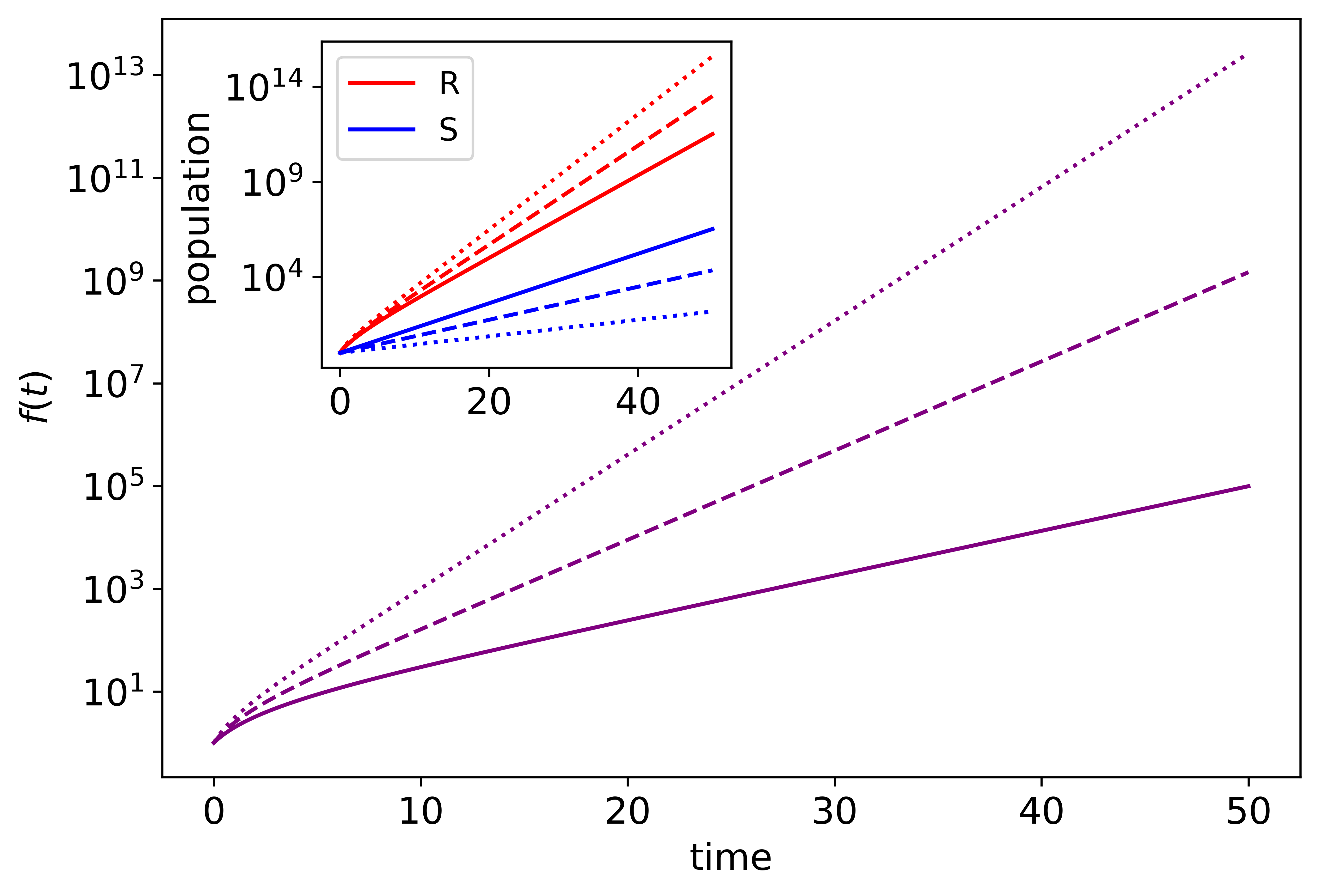}
\caption{The ratio between the two cell populations $f(t)$ for different $\varepsilon< 0$. Parameters used: $b_S=1.0, b_R = 0.8, S_0=R_0=1$ with $(\alpha_0, \delta, \varepsilon) = (0.9, 0.1, -0.6)$ for the dotted line, $(0.8, 0.2, -0.4)$ for the dashed line, and $(0.7, 0.3, -0.2)$ for the solid line. Inset: Growth curves of $S$ and $R$.}\label{fig:e_less_0}
\end{figure}

When $\varepsilon <0$, namely the effective growth rate of $R$ is greater than that of $S$, then $R$ population fixates, or $f(t) = \infty$, in the long-time limit with a characteristic time scale of $|1/\varepsilon|$, shown in Fig.\ref{fig:e_less_0}.

\begin{figure}[ht]
\centering
\includegraphics[width=0.5\textwidth]{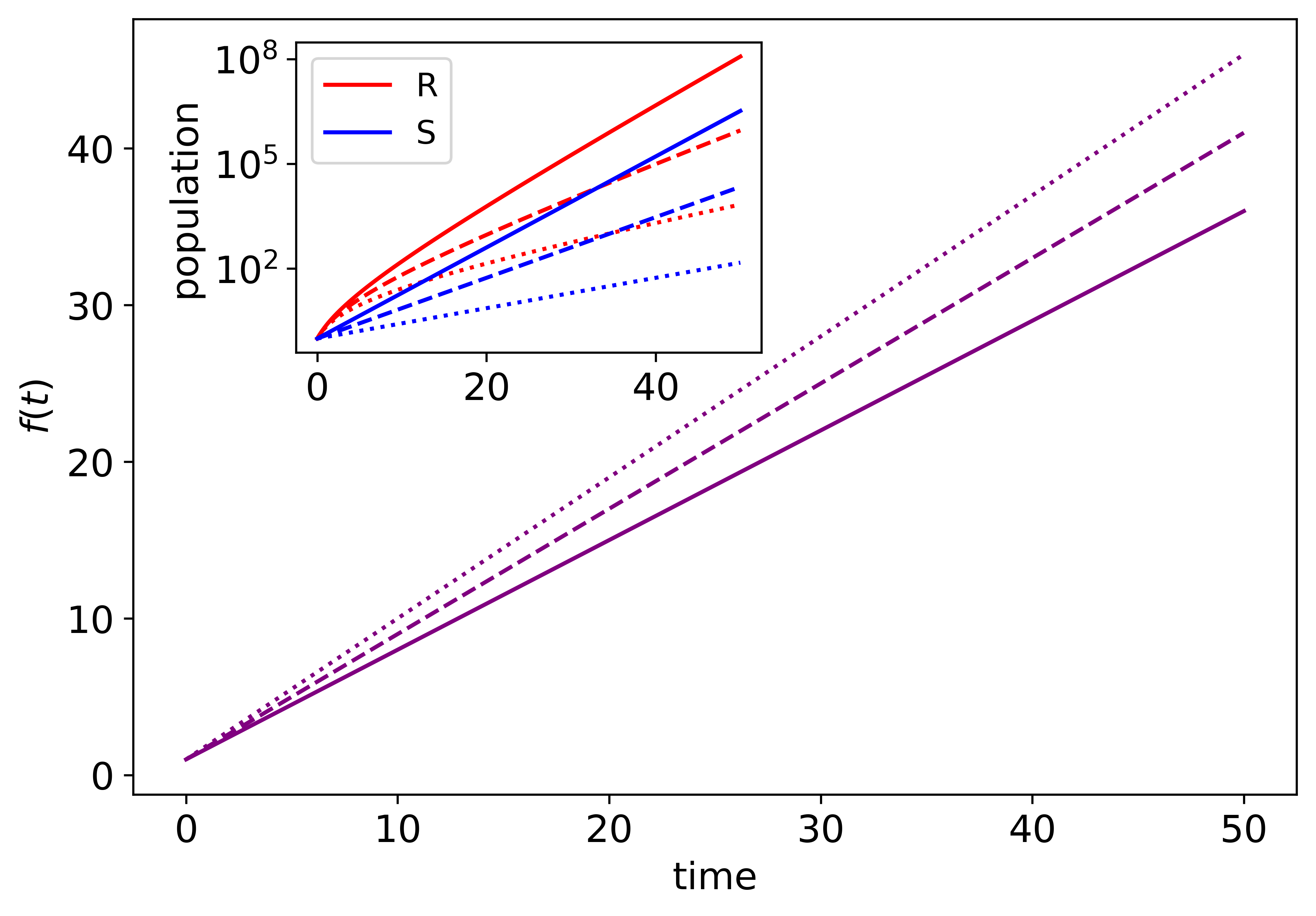}
\caption{The ratio between the two cell populations $f(t)$ for $\varepsilon=0$. Parameters used: $b_S=1.0, b_R = 0.8, S_0=R_0=1$ with $(\alpha_0, \delta) = (0.9, 0.7)$ for the dotted line, $(0.8, 0.6)$ for the dashed line, and $(0.7, 0.5)$ for the solid line. Inset: Growth curves of $S$ and $R$.}\label{fig:e_equal_0}
\end{figure}

For the case of $\varepsilon = 0$, $f(t) = R_0/S_0+\alpha_0 t$, indicating that the dominance of $R$ increases linearly with $t$, slower than the $\varepsilon <0$ case above. Still in the long-time limit, $1/f(t\rightarrow \infty) = 0$, leading to $R$-fixation. Fig.\ref{fig:e_equal_0} shows the linear increase in $f(t)$, with slopes given by $\alpha_0$. Note the different scales of $f(t)$ in Figs. \ref{fig:e_less_0} and \ref{fig:e_equal_0}.

\begin{figure}[ht]
\centering
\includegraphics[width=0.5\textwidth]{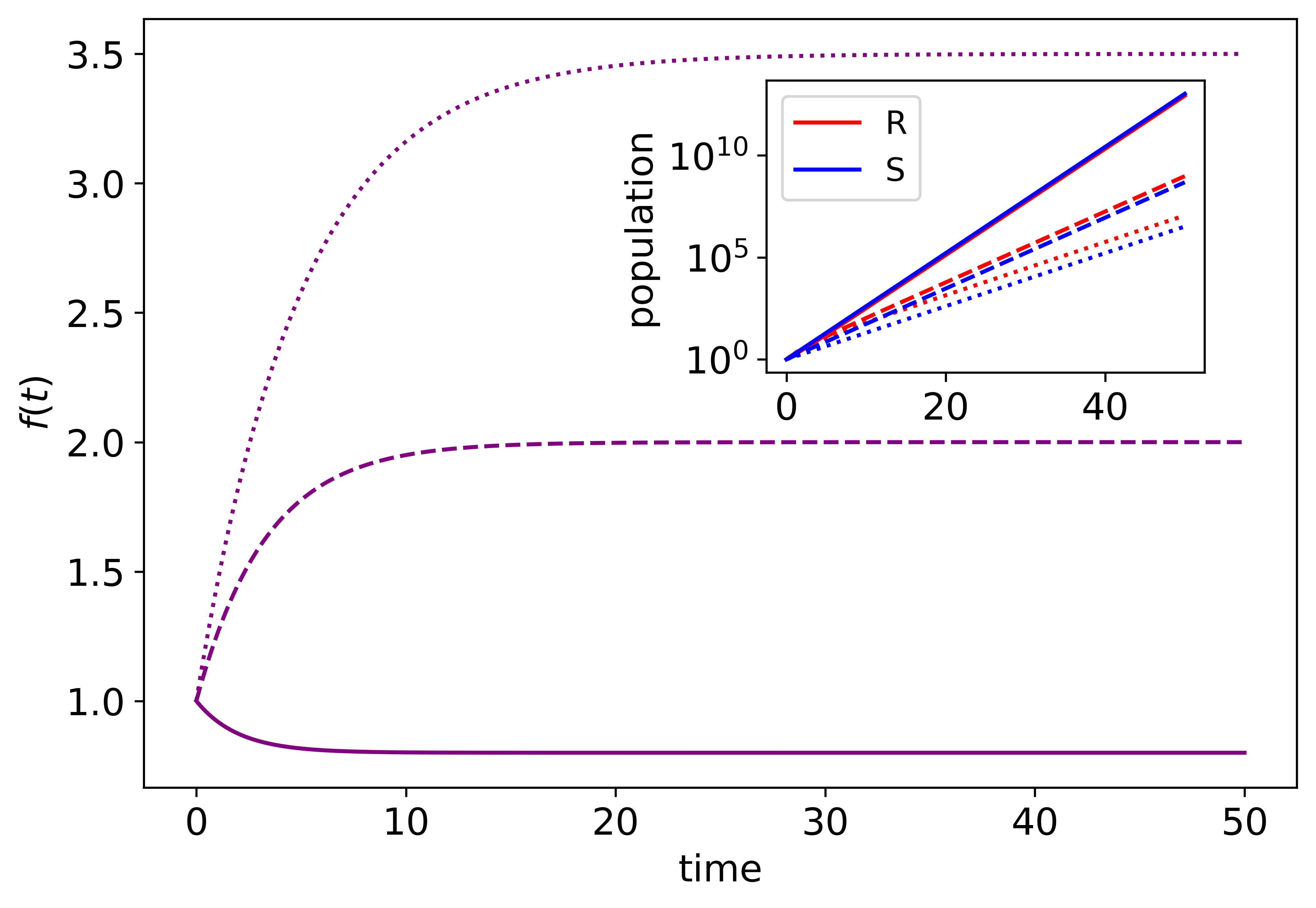}
\caption{The ratio between the two cell populations $f(t)$ for different $\varepsilon> 0$. Parameters used: $b_S=1.0, b_R = 0.8, S_0=R_0=1$ with $(\alpha_0, \delta, \varepsilon) = (0.7, 0.7, 0.2)$ for the dotted line, $(0.6, 0.7, 0.3)$ for the dashed line, and $(0.4, 0.7, 0.5)$ for the solid line. Inset: Growth curves of $S$ and $R$.}\label{fig:e_larger_0}
\end{figure}

When $\varepsilon >0$, the system displays an  $S$-$R$ coexistence as $f(t\rightarrow\infty) = \alpha_0/\varepsilon$. Either $S$ or $R$ can be the majority in the entire population depending on the competition between transformation and growth differentials: In the case of $R$-dominance, $\alpha_0>\varepsilon$. Together with the $S$-$R$ coexistence condition $\varepsilon >0$, this gives $\alpha_0\in \left((b_S-\tilde{b}_R)/2,(b_S-\tilde{b}_R)\right)$. And similarly,  $S$-dominance is reached when $\alpha_0 \in\left(0,(b_S-\tilde{b}_R)/2\right)$. In Fig.\ref{fig:e_larger_0}, we show two cases where $R$ is the majority in steady state (dotted and dashed lines), and one where $S$ dominates (solid line).

\subsection{Conjugation with constant $\gamma$\label{sec:conj}}
Having seen that transformation-only leads to either $R$-fixation or stable coexistence, we now turn to the complimentary case where $\alpha(P) =0$ and conjugation is the only HGT mechanism. Eqns.\eqref{eqn:sbirth} and \eqref{eqn:rbirth} contain a non-linear term, similar to the Lotka-Volterra type interaction \cite{lotka20,volterra26}. We find two fixed points: $(S_1^\ast, R_1^{\ast})=(0,0)$ and $(S_2^\ast, R_2^{\ast})=(-\tilde{b}_R/\gamma,b_S/\gamma)$. 

For there to be a coexistence regime where both $S$ and $R$ are non-negative, $\tilde{b}_R$ must be negative. In this case, $(S_1^\ast, R_1^{\ast})$ is a saddle point which $S$ evolving away from and $R$ into. Furthermore, from Eqns.\eqref{eqn:sbirth} and \eqref{eqn:rbirth} we find the trace and the determinant of the Jacobian matrix at $(S_2^\ast, R_2^{\ast})$ to be $0$ and $-b_S\tilde{b}_R$, respectively. Using the standard stability analysis \cite{strogatz}, we find that $(S_2^\ast, R_2^{\ast})$ is a neutrally stable center. 

In this case, the ratio $f^\ast \equiv R_2^\ast/S_2^\ast = -b_S/\tilde{b}_R$ does not depend on the conjugation $\gamma.$ What maybe slightly counterintuitive is that for $f^\ast<1$, i.e. a center with more $S$-cells than $R$, $b_S$ is constricted to be in $(0, -\tilde{b}_R)$, rather than a greater value. We will return to this point in the next section when we consider the combined effect of transformation and conjugation to the overall population.

To determine the individual closed orbits, we first eliminate $t$ by dividing $dS/dt$ by $dR/dt$ and then separate the variables:
\begin{eqnarray*}
(\tilde{b}_R\frac{1}{S}+\gamma)dS= (b_S\frac{1}{R}-\gamma)dR&&\\
(\tilde{b}_R\ln S +\gamma S)-(b_S\ln R- \gamma R) &=&C,
\end{eqnarray*}
where $C$ is the constant of integration. We now have an expression to describe the trajectory for the conjugation-only system:
\begin{equation}
R^{b_S}S^{-\tilde{b}_R}e^{-\gamma(R+S)}=e^{C}\label{eq:orbit}
\end{equation}

\begin{figure}[ht]
\begin{tabular}{cc}
\includegraphics[width=0.23\textwidth]{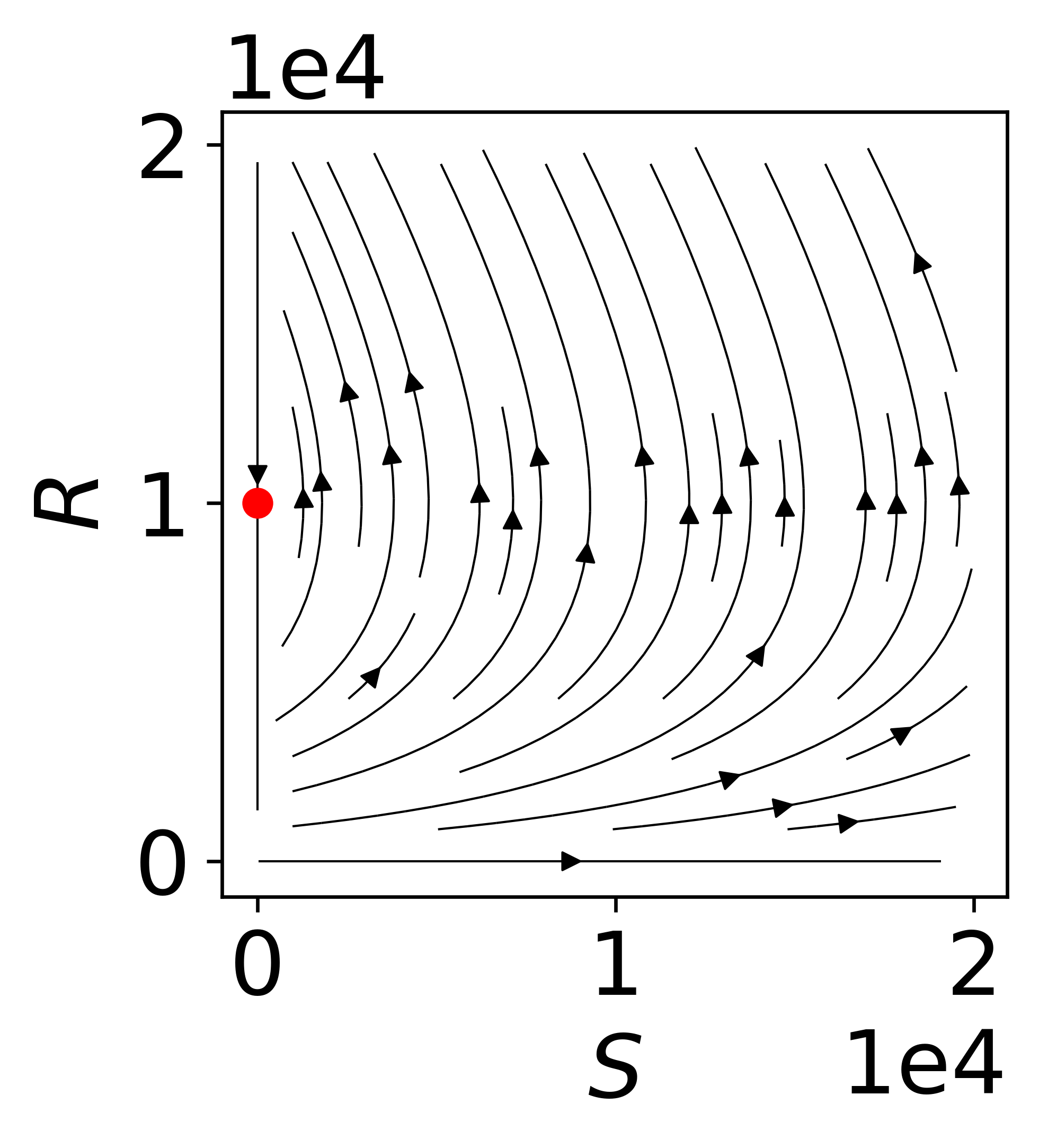}
\includegraphics[width=0.23\textwidth]{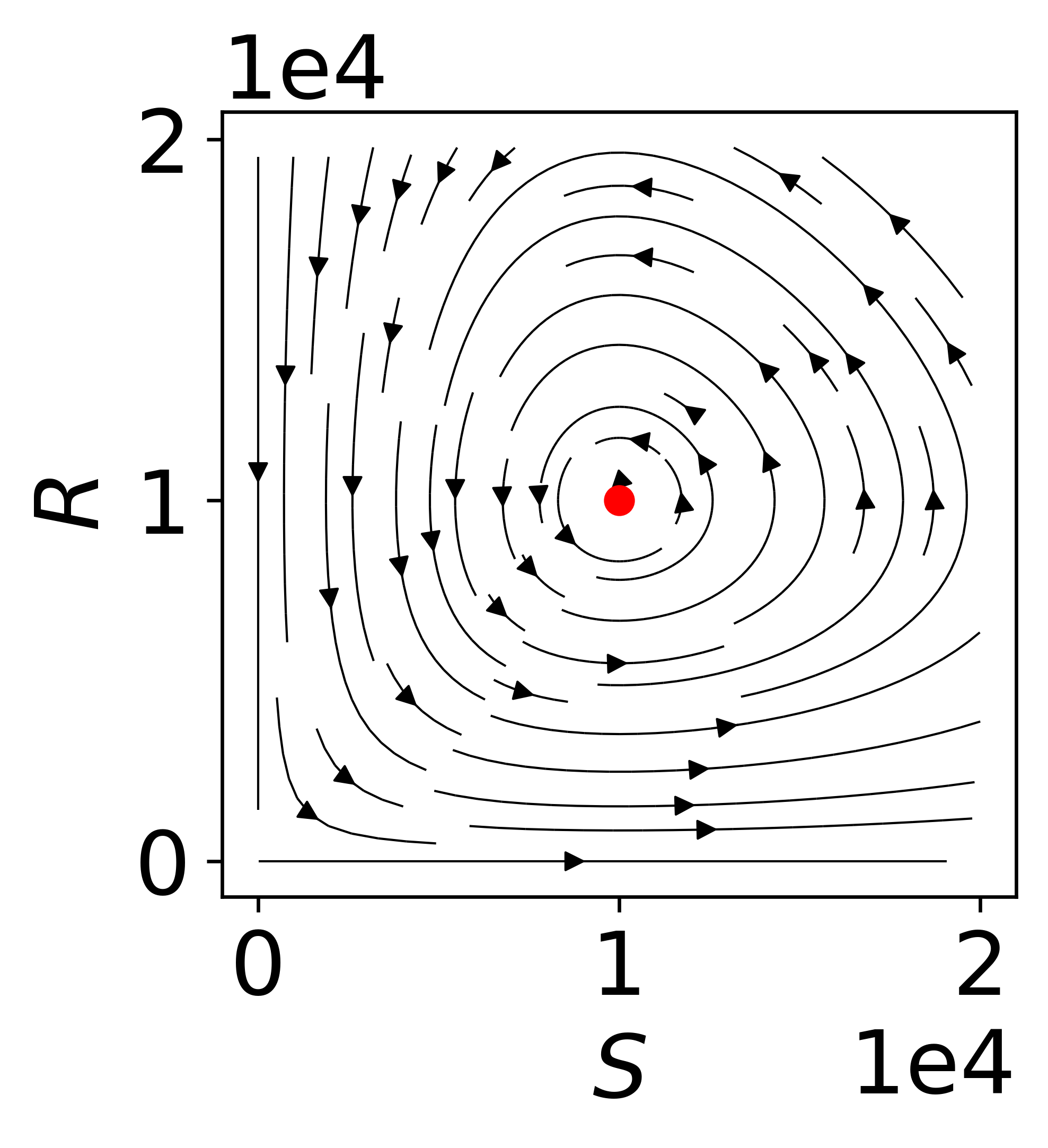}
\end{tabular}
\caption{Phase portraits for $S$ and $R$. $b_S =1, \gamma =10^{-4}$ and $-\tilde{b}_R/\gamma= 1$ (left) and $10^4$ (right). The red dot indicates the center of the orbits $(S_2^\ast, R_2^{\ast})$.}\label{fig:orbit}
\end{figure}

We show in Fig.\ref{fig:orbit} the phase portraits of the two sub-populations for $\tilde{b}_R<0.$ Setting $b_S =1$ and varying the ratio between $-\tilde{b}_R$ and $\gamma$ using the range of conjugation rates in Table~\ref{table:rxn}, we see the competition between growth and conjugation leads to different trajectories, all around the center $(S_2^\ast, R_2^{\ast})$, indicated as a red dot. 

\begin{figure}[H]
\centering
\includegraphics[width=.5\textwidth]{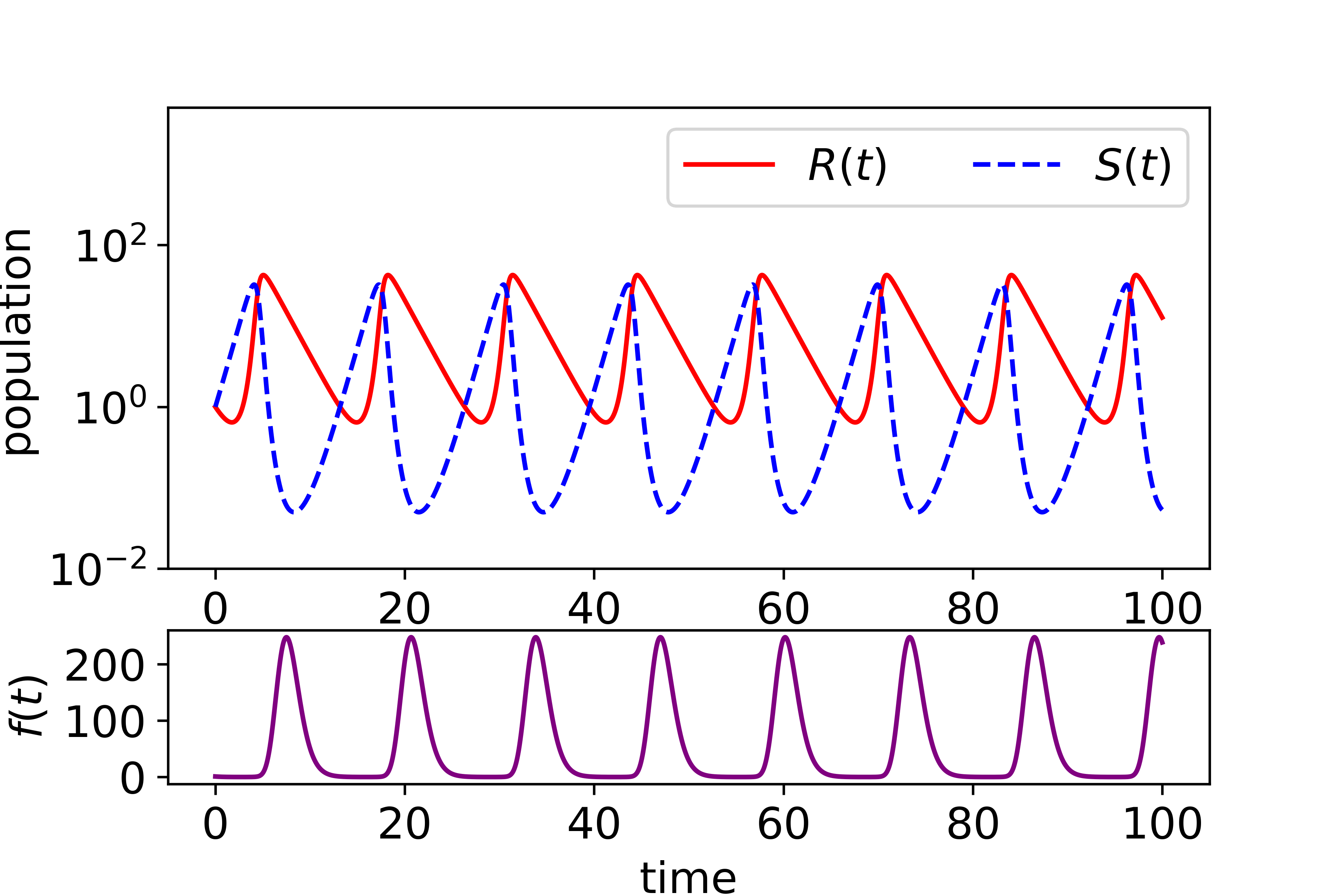}
\caption{Top: Oscillations of the $S$ and $R$ population (color online) with conjugation only. Bottom: The corresponding oscillations in the sub-population ratio $f(t)$. Parameters: $S_0=R_0=1, \gamma =0.1, b_S = 1.0$ and $\tilde{b}_R = -0.5$.}\label{fig:orbit_osc}
\end{figure}

In Fig.\ref{fig:orbit_osc} we observe the oscillations in $S$ and $R$ on one of the closed orbits. The fact that $S$-$R$ can coexist when $\tilde{b}_R<0$ is particularly interesting because the increase in $R$-cells is through converting more $S$-cells by conjugation. A closer comparison between conjugation-only and conjugation with transformation is provided in the following section.

\section{Combined effects of transformation and conjugation\label{sec:sim}}
Informed by the above discussions on the individual effects of transformation and conjugation, we study the model of which the full dynamics is described in Eqns.\eqref{eqn:sbirth}-\eqref{eqn:pfree} with a range of parameters to chart the different regimes of the sub-populations of $S$ and $R$ cells, with special attention on the $S$-$R$ coexistence conditions.

From the discussions in Section \ref{sec:conj}, we learned that a physical coexistence of $S$ and $R$ cells requires $b_R<\delta$ in the conjugation-only case. With transformation included, first let's consider the case of constant transformation, or $K_P \ll P$. Thus $\alpha(P)\rightarrow \alpha_0$ and the new non-trivial fixed points are:
\begin{align}
(S^\ast, R^\ast) = \left((\frac{\alpha_0}{b_S}-1)\cdot\dfrac{\tilde{b}_R}{\gamma}, \dfrac{b_S-\alpha_0}{\gamma}\right) \equiv \left(-\dfrac{\tilde{b}_S\tilde{b}_R}{\gamma b_S}, \dfrac{\tilde{b}_S}{\gamma}\right) ,\label{eq:fp_ct}
\end{align} 
with $f^\ast =- b_S/\tilde{b}_R$. The fixed point at $(0, 0)$ remains a saddle point. It is worth noting that the value $f^\ast$ is the same as in the conjugation-only case. Meanwhile, the overall population of the system approaches $\dfrac{\tilde{b}_S}{\gamma}\left(1-\dfrac{\tilde{b}_R}{b_S}\right)$. 

Combining transformation and conjugation brings new behaviors in the overall system. Unlike the conjugation-only case where $S$ and $R$ evolve along a closed orbit centered around $(-\tilde{b}_R/\gamma,b_S/\gamma)$ with temporal oscillations, as shown in Figs.\ref{fig:orbit}-\ref{fig:orbit_osc}, once transformation is introduced, these orbits turn into spirals. We find an $S$-$R$ coexistence when $\tilde{b}_R<0$ and $\tilde{b}_S>0$. To determine the stability of $(S^\ast, R^\ast)$, we can again examine the Jacobian at this point: the trace and the determinant is $\tilde{b}_R\dfrac{\alpha}{b_S}<0$ and $-\tilde{b}_S\tilde{b}_R >0$. Thus $(S^\ast, R^\ast)$ will be a stable node or a stable spiral. However, for it to be a stable node, $\tilde{b}_R> -4(1+\tilde{b}_S/\alpha)^2\tilde{b}_S$. This means when $\tilde{b}_R$ is almost one order of magnitude greater than $\tilde{b}_S$, $(S^\ast, R^\ast)$ is a stable node. In other cases where $\tilde{b}_R\sim \tilde{b}_S$, the system has a stable spiral towards $(S^\ast, R^\ast)$.

\begin{figure}[H]
\centering
\includegraphics[width=.5\textwidth]{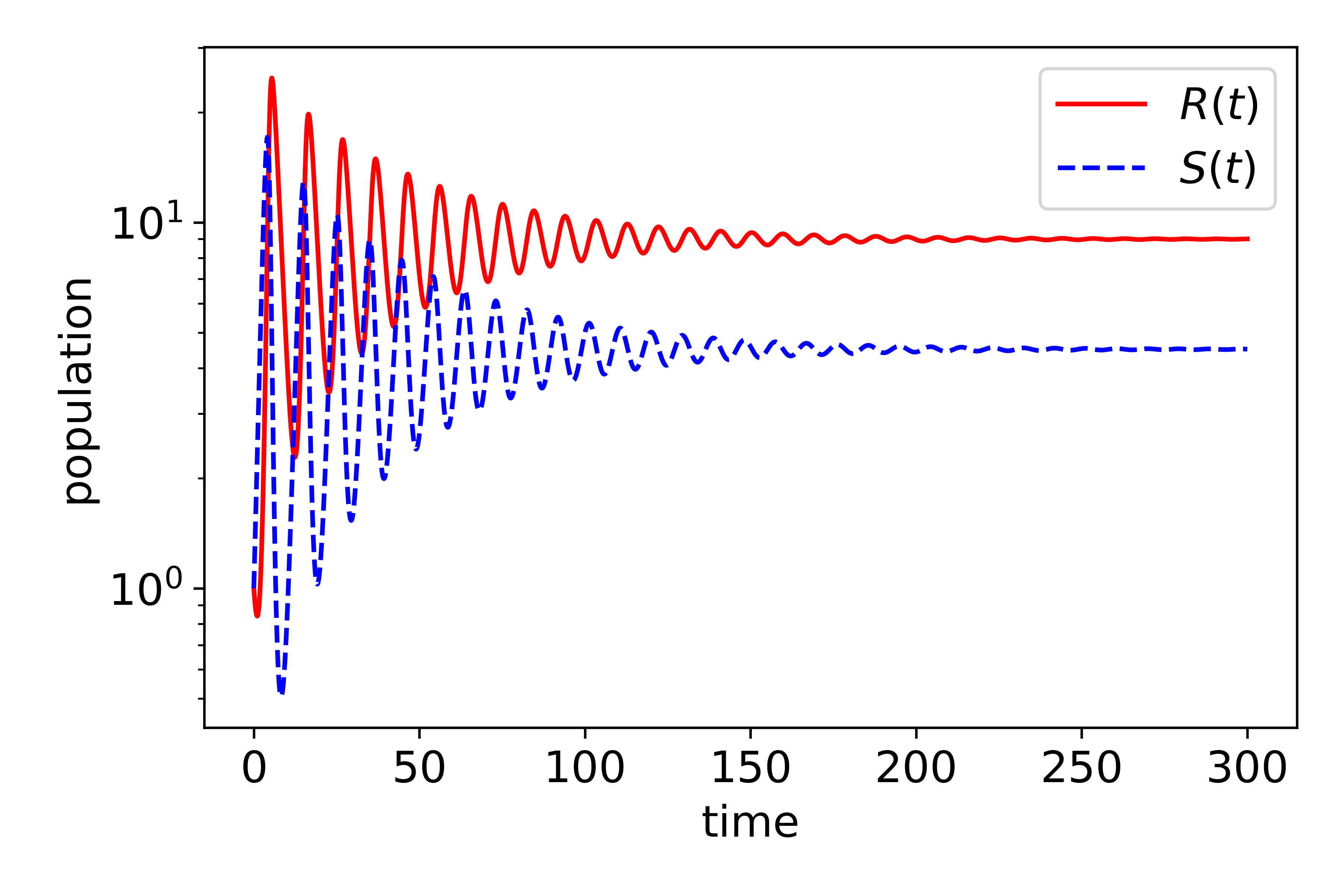}
\caption{Timetrace of the $S$ and $R$ population with $S_0=R_0=1, P_0 =1, K_P=0.01P_0, \alpha_0=0.1, \gamma =0.1, b_S = 1.0$ and $\tilde{b}_R = -0.5$.}\label{fig:RSt_spiral}
\end{figure}
\begin{figure}[H]
\centering
\includegraphics[width=.5\textwidth]{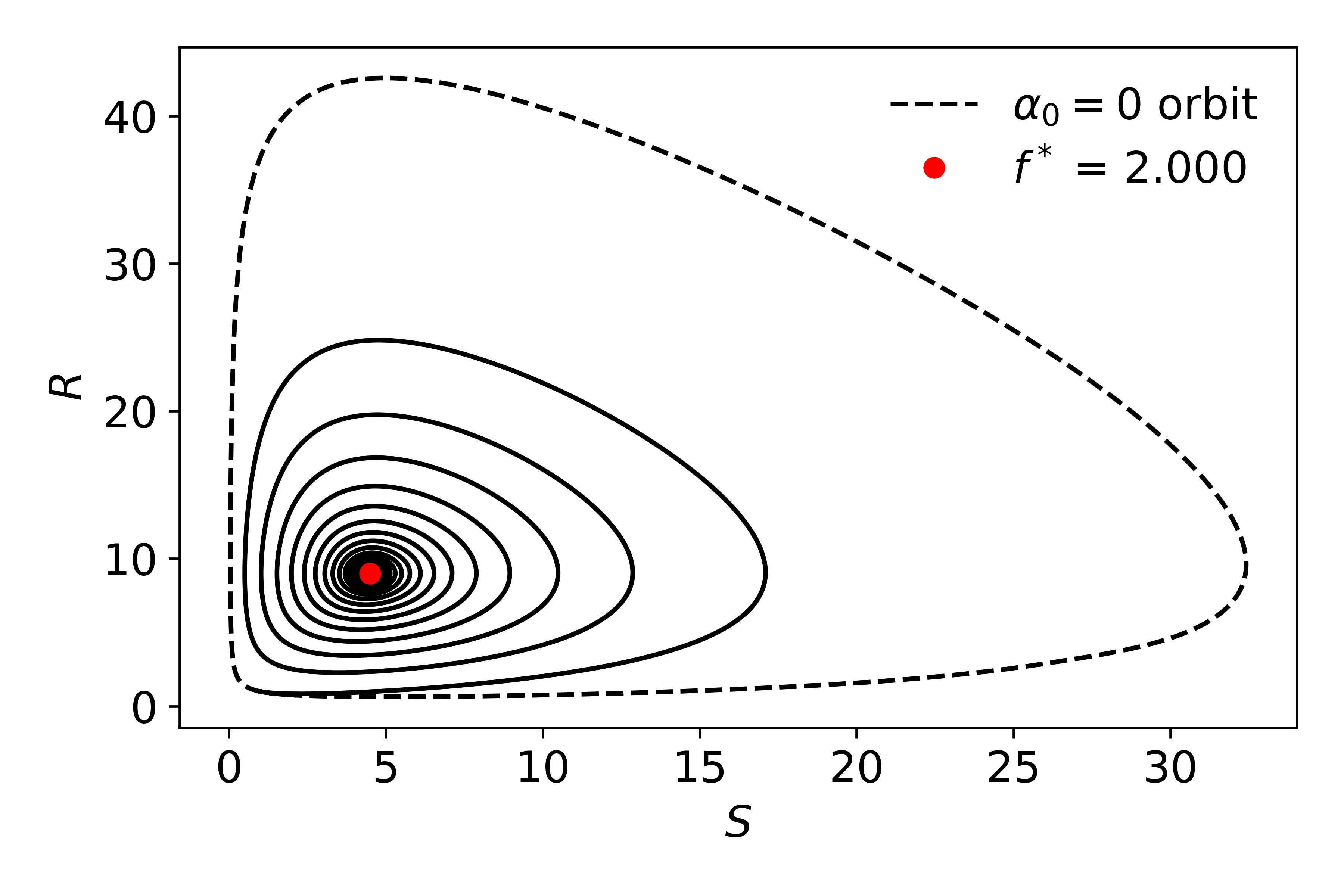}
\caption{The trajectory of $S$-$R$ spirals towards $(S^\ast, R^\ast)$ with $S_0=R_0=1, P_0 =1, K_P=0.01P_0, \alpha_0=0.1, \gamma =0.1, b_S = 1.0$ and $\tilde{b}_R = -0.5$, and reaches $f^\ast = 2.0$. The dashed line shows the stable orbit when $\alpha_0 =0$ as a reference. }\label{fig:RS_spiral}
\end{figure}

In Fig.\ref{fig:RSt_spiral}, the system starts with $S_0=R_0 =1$ and the initial oscillations of the two sub-populations quickly settles into $(S^\ast, R^\ast) = (4.5, 9)$, as given in Eq.\eqref{eq:fp_ct}. Their trajectory is shown in Fig.\ref{fig:RS_spiral}, with the steady state being the red dot at $(S^\ast, R^\ast)$. We also show the closed orbit for the same set of parameters except for $\alpha_0=0$ as a reference. As predicted, the $S$-$R$ coexists with a stable ratio of $f^\ast = 2.0$ in the long-time limit. Noticeably, the spiral is smaller than the conjugation-only orbit due to the damped oscillations in $S$ and $R$ when transformation is introduced ($\alpha_0 = 0.1$ in this case). The general shape of the spiral will depend on the initial conditions of $S_0$ and $R_0$, which does not affect the final coexistence ratio $f^\ast$.

\begin{figure}[H]
\centering
\includegraphics[width=.5\textwidth]{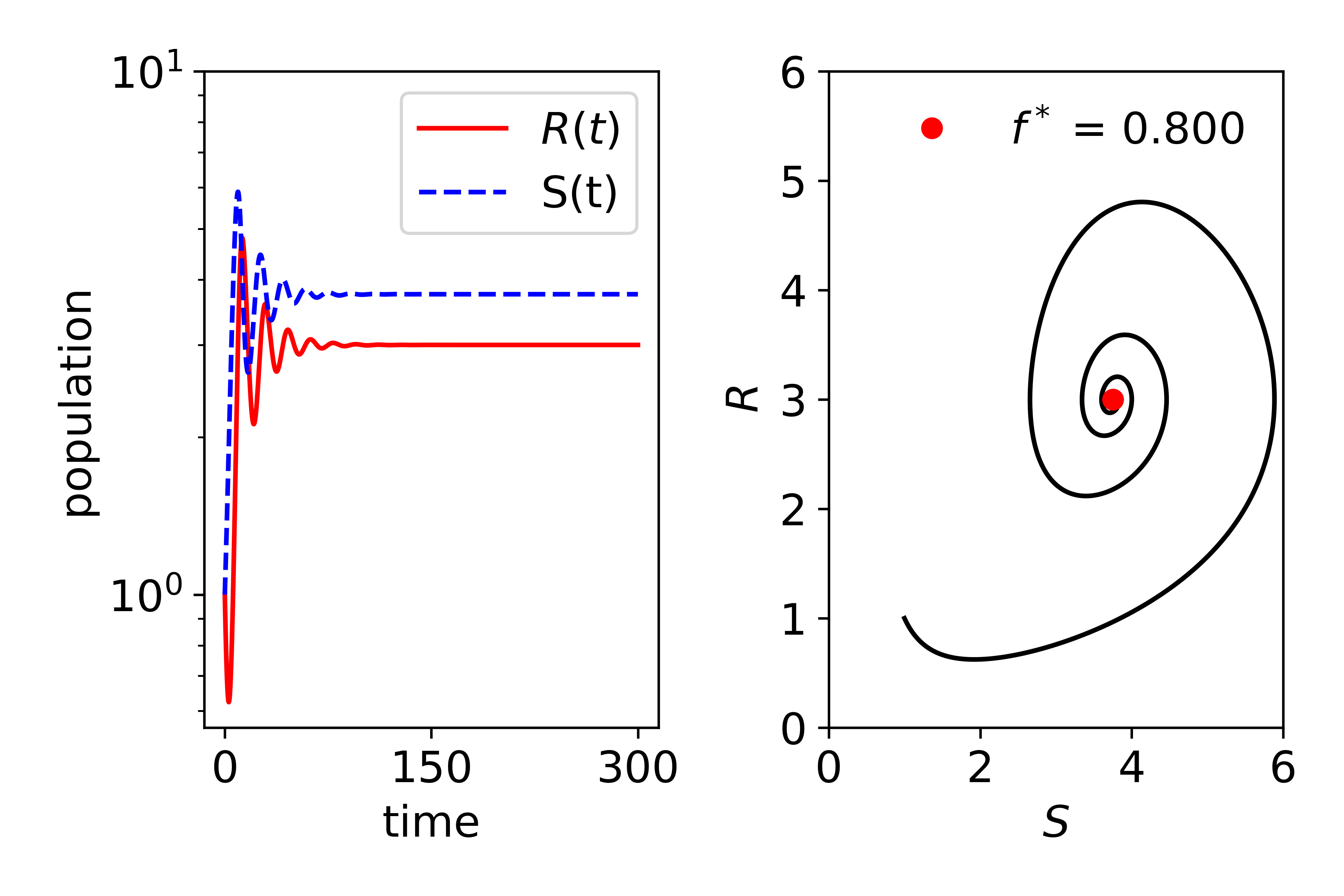}
\caption{Left: Timetrace of the $S$ and $R$ population with $S_0=R_0=1, P_0 =1, K_P=0.01P_0, \alpha_0=0.1, \gamma =0.1, b_S = 0.4$ and $\tilde{b}_R = -0.5$. Right: The trajectory of $S$-$R$ spirals towards $(S^\ast, R^\ast)$ with $f^\ast =0.8$. }\label{fig:RS_spiral2}
\end{figure}

As alluded to in Section \ref{sec:conj}, for $S$ to be the majority in the overall population in steady state, the growth rate $b_S$ needs to be less than $-\tilde{b}_R$. For example, when $b_S =0.4$ while $-\tilde{b}_R = 0.5$, as in Fig.\ref{fig:RS_spiral2}, the final $S$-$R$ coexistence ratio $f^\ast =0.8$. When $b_S<-\tilde{b}_R$, even though $S$ is doubling at a lower rate, that also leads to fewer $S$ being conjugated or transformed into $R$. The final coexistence state contains more $S$ than $R$ cells.

Next let's turn to the case where transformation rate is no longer a constant. Here two parameters are at play: the initial concentration of plasmids in the environment, $P_0$, and the transformation kinetics $K_P$. The former obviously depends on the condition of the growth medium, while the latter is a biochemical parameter where a higher $K_P$ means a lower plasmid-bacteria affinity. The dynamics of the plasmids in this case is given in Eq.\eqref{eqn:pfree}.

\begin{figure}[H]
\centering
\includegraphics[width=.5\textwidth]{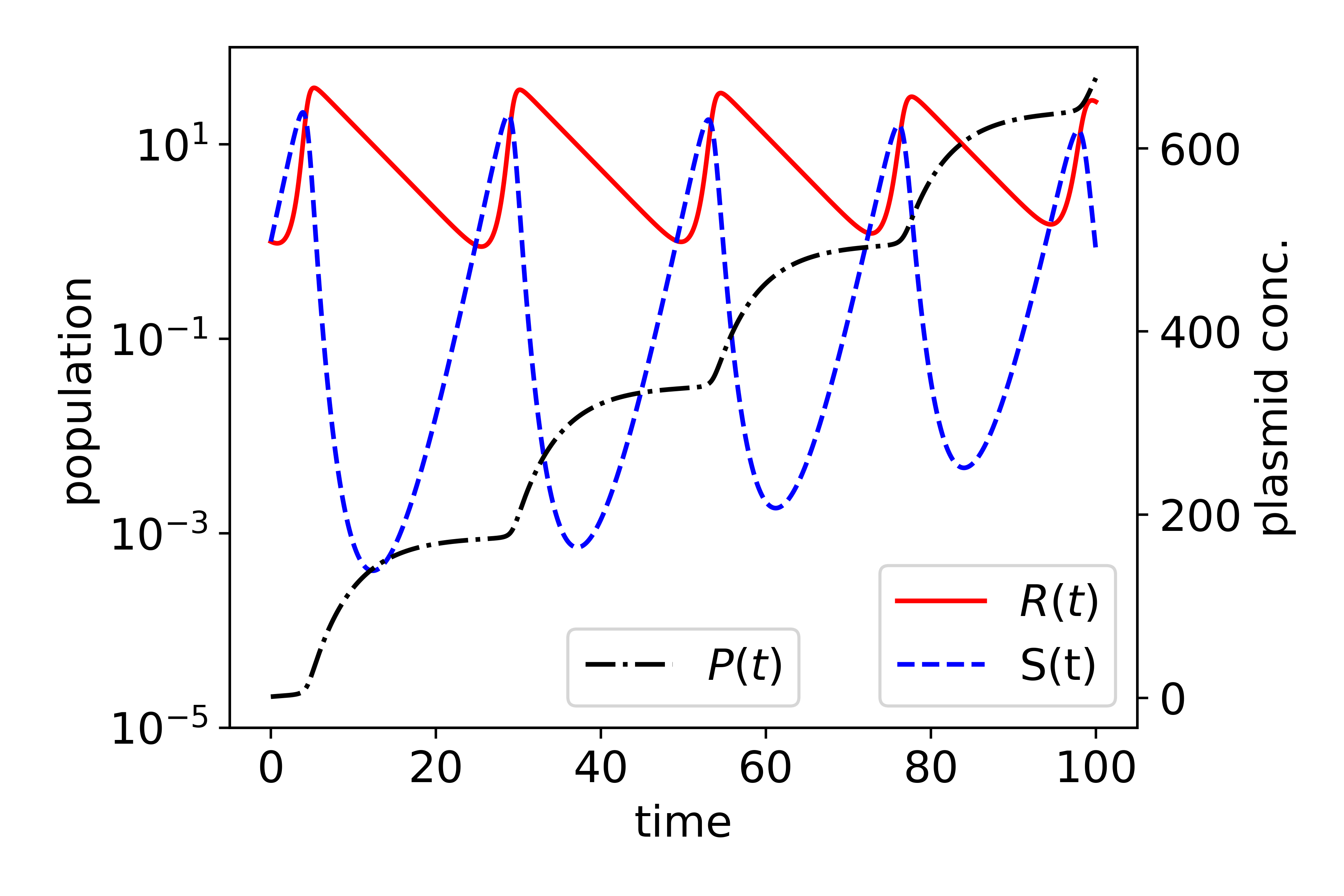}
\caption{Timetrace of $S$ and $R$, along with plasmid concentration $P$ with its scale to the right. Parameters used: $S_0=R_0=1, P_0 =1, K_P=10^3P_0, \alpha_0=0.1, \gamma =0.1, b_S = 1.0$ and $\tilde{b}_R = -0.2$.}\label{fig:R_comp}
\end{figure}

If the lysis rate $\delta$ is sufficiently large, or $R$ is the dominating species, then the plasmid concentration will eventually build up and it becomes similar to the $\alpha(P)= \alpha_0$ case as discussed above. However, the transient behavior in this regime contains some interesting details. For an environment with few low-affinity plasmids, $K_P\gg P$, the transformation rate is:
\[\alpha(P)=\alpha_0 \dfrac{P}{P+K_P} \approx \alpha_0 \dfrac{P}{K_P}.\]
The stability condition with conjugation-only case also requires $\delta > b_R$ while transformation with the low-affinity plasmids is slow. The buildup of plasmids -- consequently $\alpha(P)$ -- therefore sees a ``burst''-like behavior, shown in Fig.\ref{fig:R_comp}, where we look at low-affinity plasmids with $K_P = 10^3 P_0$. The bursts coincide with the decrease of $R$ due to lysis and increase the fastest when $S$ reaches its minimum so that $dP/dt$ is the greatest. Additionally, the burst cycle tracks the $S$-$R$ oscillations.

\section{Summary and Discussion\label{sec:sum}}
We studied a two-species population model aimed to explore the interplay between bacterial transformation and conjugation in addition to their natural growth cycles. Using a combination of theoretical and numerical analysis, we analyzed the complete effects of transformation: the overall population will either be fixated by $R$-cells or in stable co-existence. In the latter case, the majority cell type is determined by the transformation rate $\alpha_0$ and $\varepsilon$, the difference between the effective growth rates of the two sub-populations. The conjugation-only case gives $R$-fixation, unless when the effective growth rate of $R$-cells $\tilde{b}_R$ is less than $0$, in which case we can actually observe a stable orbit for $S$ and $R$, with each sub-population oscillating. This could potentially be interesting in the biologically-relevant context: carrying an additional plasmid confers growth disadvantage for $R$-cells as $\tilde{b}_R<0$, they can nevertheless survive steadily only by conjugating more $S$ to $R$. More interestingly, when transformation and conjugation are both considered, the orbit seen in the conjugation-only case turns into a spiral with a stable center, indicating a steady state where the final ratio between the two sub-populations $f^\ast$ reaches a constant.

In our numerical analysis of the model, the reaction rates tested are not directly comparable to the experimentally obtained ones listed in Table \ref{table:rxn} because the equations of the system were first non-dimensionalized. However, our results point to the important regimes where interesting behaviors, such as stable co-existence and oscillation, emerge when the {\it relative} reaction rates are taken into consideration. In particular, the difference between the two effective growth rates $\varepsilon$ in the case of transformation, and the ratio between the two effective growth rates characterized by $f^\ast = -b_S/\tilde{b}_R$ in the case of conjugation are direct indication on the overall population behavior.

From a modeling perspective, our findings are novel in that we quantitatively analyze the effects of two main horizontal gene transfer (HGT) mechanisms at a population level. The results are significant especially with the use of a minimal model with few parameters, mostly experimentally accessible. The lysis rate $\delta$ is the only estimated parameter here, due to its elusiveness because cell lysis often involves various other environmental and physiological factors. And yet it is crucial in determining whether the system has a steady state, or which sub-population dominates. We thus continue to seek experimental breakthroughs on further clarification on the lysis process. 

The results from this study are potentially helpful to a wider range of explorations. In our study, there is no limit on the population growth aside from the dynamical parameters. It is conceivable to impose an environmental constraint such as a carrying capacity and analyze the difference. For this study, the effects of antibiotic resistance is binary, in the form of carrying the plasmid, and deterministic. It is possible for cells to lose a plasmid without lysing (e.g. \cite{boe1987}), in which case there needs to be an additional reaction to capture $R\longrightarrow S+P$ process. And in terms of stochasticity, our preliminary study using kinetic Monte Carlo shows consistent results with what we present here, and is more powerful to generalize to more reactions with other complications. Further extensions to include the population spatial structure beyond the existing well-mixed population are also being considered. The spatial aspect is especially compelling to investigate because both transformation and conjugation processes depend on local availability of free plasmids as well as direct contacts between $S$ and $R$ cells. Thus the diffusion of plasmids as well as the cells are expected to be relevant, an aspect not necessary for the well-mixed case.

\begin{acknowledgements}
The authors acknowledge the financial support from National Science Foundation grants DMR-1248387 and DMR-1702321. JJD is grateful for the hospitality of Dr.~Stefan Klumpp at the University of G{\"o}ttingen where the initial stage of this research was carried out. The manuscript benefitted from fruitful discussions with M.D. Eichenlaub.
\end{acknowledgements}

\bibliographystyle{unsrt}
\bibliography{ref_conjtrans}

\end{document}